\documentclass{article}
\usepackage{spconf,amsmath,graphicx}

    \usepackage{amsfonts}
    \usepackage{mathtools}
    \usepackage{amssymb}
    \usepackage{stmaryrd}
    \usepackage{gensymb}
    \usepackage{hyperref}
    \usepackage{color}
    \usepackage{tikz}
    \usepackage{bbold}
    \usepackage{pgfplots}
    \usepackage{diagbox}

    \usepackage{hyperref}
    \newcommand{\figref}[1]{Figure~\ref{#1}}
    
    \newcommand{\secref}[1]{Section~\ref{#1}}
    \renewcommand{\eqref}[1]{Equation~\ref{#1}}
    
    \newcommand{\prref}[1]{Property~\ref{#1}}

    \newcommand{\ie}{\emph{i.e}}
    \newcommand{\st}{\text{s.t.}}
    \newcommand{\eg}{\emph{eg.}}
    \newcommand{\etal}{\emph{et al.}}
    
    \newcommand{\setR}{\text{$\mathbb{R}$}}
    \newcommand{\vecOne}[1]{\text{$\mathbb{1}_{#1}$}}
    \newcommand{\vecZero}[1]{\text{$\mathbb{0}_{#1}$}}
    
    \newcommand{\pair}[2]{(#1; #2)}
    \newcommand{\intInterval}[2]{\{#1; \ldots; #2\}}
    \newcommand{\interval}[2]{[#1; #2]}
    \newcommand{\set}[1]{\{#1\}}
    \newcommand{\matij}[3]{\text{$#1_{#2; #3}$}}
    \newcommand{\veci}[2]{\text{$#1_{#2}$}}
    \newcommand{\dist}{\text{$d$}}
    \renewcommand{\~}[1]{\text{$\widetilde{#1}$}}
    \newcommand{\scalar}[2]{\text{$\langle #1; #2 \rangle$}}
    
    \newcommand{\G}{\text{$\mathcal{G}$}}
    \newcommand{\V}{\text{$\mathcal{V}$}}
    \newcommand{\E}{\text{$\mathcal{E}$}}
    \newcommand{\A}{\textbf{A}}
    \newcommand{\B}{\textbf{B}}
    \newcommand{\C}{\textbf{C}}
    \newcommand{\D}{\textbf{D}}
    \newcommand{\W}{\textbf{W}}
    \newcommand{\M}{\textbf{M}}
    \newcommand{\N}{\text{$N$}}
    \renewcommand{\u}{\text{$u$}}
    \newcommand{\uc}{\text{$u_c$}}
    \renewcommand{\v}{\text{$v$}}
    
    \newcommand{\x}{\textbf{x}}
    \newcommand{\y}{\textbf{y}}
    \newcommand{\sx}{\textbf{$\hat{\x}$}}
    \newcommand{\NL}{\textbf{\L}}
    \newcommand{\I}[1]{\text{$\textbf{I}_{#1}$}}
    \renewcommand{\P}{\textbf{P}}
    \newcommand{\eigval}[1]{\text{$\boldsymbol{\Lambda}_{#1}$}}
    \newcommand{\eigvec}[1]{\text{$\boldsymbol{\mathcal{X}}_{#1}$}}
    \newcommand{\el}{\text{$\lambda$}}
    \newcommand{\ev}{\text{$\boldsymbol{\chi}$}}
    \newcommand{\sqTS}{\text{$\Delta_t^2$}}
    \newcommand{\sqFS}{\text{$\Delta_\omega^2$}}
    \newcommand{\sqGS}{\text{$\Delta_{\G;\uc}^2$}}
    \newcommand{\sqSS}{\text{$\Delta_s^2$}}
    \newcommand{\lTwo}{\text{$\ell_2$}}
    \newcommand{\uCurve}{\text{$\gamma_\uc$}}
    
    \usepackage{subcaption}
    \usepackage{amsthm}
    \newtheorem{definition}{Definition}
    \newtheorem{property}{Property}
    \newenvironment{subFigure}[1]{\hfill\begin{subfigure}{#1}\centering}{\end{subfigure}\hfill}

    \usepackage{pgfplots}
    \usepackage{tikz}
    \usepackage{tkz-graph}
    \usetikzlibrary{shapes, arrows, graphs}
    \newcommand{\colorbar}[3]{\begin{tikzpicture}\begin{axis}[hide axis, scale only axis, height=0pt, width=0pt, colorbar, point meta min=#1, point meta max=#2, colorbar style={height=#3, ytick={#1, ..., #2}}]\end{axis}\end{tikzpicture}}

\title{TOWARDS A CHARACTERIZATION OF THE UNCERTAINTY CURVE FOR GRAPHS}
\name{Bastien Pasdeloup$^{\star}$ \qquad Vincent Gripon$^{\star}$ \qquad Gr\'egoire Mercier$^{\star}$ \qquad Dominique Pastor$^{\star}$\thanks{This work was supported by the European Research Council under the European Union's Seventh Framework Programme (FP7/2007-2013) / ERC grant agreement n°~290901.}}

\address{$^{\star}$ Telecom Bretagne, UMR CNRS Lab-STICC}

\begin{document}

\maketitle

    
    \begin{abstract}
        Signal processing on graphs is a recent research domain that aims at generalizing classical tools in signal processing, in order to analyze signals evolving on complex domains.
        Such domains are represented by graphs, for which one can compute a particular matrix, called the normalized Laplacian.
        It was shown that the eigenvalues of this Laplacian correspond to the frequencies of the Fourier domain in classical signal processing.
        Therefore, the frequency domain is not the same for every support graph.
        A consequence of this is that there is no non-trivial generalization of Heisenberg's uncertainty principle, that states that a signal cannot be fully localized both in the time domain and in the frequency domain.
        A way to generalize this principle, introduced by Agaskar and Lu, consists in determining a curve that represents a lower bound on the compromise between precision in the graph domain and precision in the spectral domain.
        The aim of this paper is to propose a characterization of the signals achieving this curve, for a larger class of graphs than the one studied by Agaskar and Lu.
    \end{abstract}




    \section{Introduction}
    \label{intro}
        
        In the field of signal processing on graphs, a signal can be seen as a temporal series, associating an intensity to every observed moment.
        In this context, the support of information is unidimensional, and is represented by the axis of time.
        One of the main objectives of signal processing on graphs is to extend tools from classical signal processing to new signals, associated with more complex topologies that are represented by graphs.
        The portage of tools such as convolution, translation of a signal, or Fourier transform \cite{Shuman2013} was made possible thanks to the correspondence between frequencies in classical Fourier analysis and the eigenvalues of a certain matrix, associated with the graph.
        A signal can therefore be seen as a vector associating an intensity to every node in the graph, and having a spectral decomposition according to the eigenvectors of a particular matrix that is dependent on the graph.
        
        With the aim of porting tools from classical signal processing to signal processing on graphs, Agaskar and Lu \cite{Agaskar2012, Agaskar2012b} proposed an adaptation of Heisenberg's uncertainty principle \cite{Folland1997}.
        This principle classically states that a signal cannot be fully localized both in the time and frequency domains.
        The authors have shown that such a compromise also exists in signal processing on graphs.
        Therefore, a signal on a graph cannot be fully localized both in the graph and spectral domains.
        
        There are alternative definitions for the uncertainty principle on graphs (\eg{} \cite{Tsitsvero2015}).
        The choice of a definition is a debated topic in the community of signal processing on graphs.
        The approach of Agaskar and Lu has the advantage to give a lot of importance to the underlying graph in the computation of the spread of a signal, what we believe should be an prominent factor.
        
        In more details, Agaskar and Lu have shown that the compromise between spectral precision and graph domain precision is dependent of the graph holding the signals.
        They have introduced a notion of uncertainty curves representing, for a given graph and a chosen node, the pairs \pair{graph domain precision}{spectral precision} being Pareto optima.
        Moreover, they have shown that every uncertainty curve is convex, and have illustrated for a particular example that the curve could be described by a portion of an ellipse.
        
        In this paper, we extend the work of Agaskar and Lu, by proposing a method to characterize the signals that reach the uncertainty curve of a given graph.
        This method allows one to reduce the search space of signals reaching the uncertainty curve, \ie{} being a Pareto optimum in terms of spectral and graph domain precisions.
        
        This document is organized as follows: in \secref{spog}, we present in more details the notions from signal processing on graphs theory that are required for a full insight of our work.
        in \secref{uncertainty}, we detail the notion of uncertainty on graphs, as introduced in \cite{Agaskar2012}.
        Finally, in \secref{contribution}, we extend the results of Agaskar and Lu, and propose a method to characterize the signals reaching the uncertainty curve.
        \secref{conclu} concludes this document, and proposes extensions to our work.
        
    %


    \section{Signal processing on graph}
    \label{spog}
        
        In the field of signal processing on graphs, the support for signals is not only the time, but can be a more complex structure.
        To represent such a support, we introduce the notion of graph:

        \begin{definition}[Graph]
            A graph (simple, non-directed) \G{} is a tuple $(\V, \E, \W)$, where $\V = \intInterval{1}{\N}$ is a set of \N{} nodes, $\E = \V \times \V$ is a set of edges, and \W{} is a matrix representing the weights associated with the edges: $\forall \u, \v \in \V : \pair{\u}{\v} \in \E \Leftrightarrow \matij{\W}{\u}{\v} = \matij{\W}{\v}{\u} \neq 0$.
        \end{definition}

        A signal on a graph is a vector associating an intensity to every node in the graph:
        
        \begin{definition}[Signal on graph]
            A signal $\x = \set{\x_1; \dots; \x_\N}$ on a graph \G{} of \N{} nodes is a vector in $\setR^\N$.
            Without loss of generality, we study in this document signals that have been normalized, \ie{} such as $\|\x\|_2 = 1$.
        \end{definition}
        
        \figref{graphWithSignal} illustrates a signal \x{} on a graph with a star topology.
        The values of the components of \x{} are represented using colors:
        
        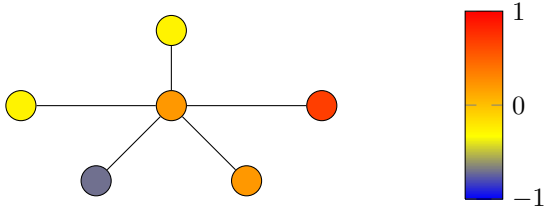
\begin{figure}[h!]
            \begin{subFigure}{0.7\linewidth}

    \tikzstyle{graphNode} =
    [
        circle,
        draw,
        text centered,
        minimum height=0.4cm
    ]
    
    \tikzstyle{graphEdge} =
    [
        draw
    ]
    
    \definecolor{signalValue1}{RGB}{255, 62, 0}
    \definecolor{signalValue2}{RGB}{255, 152, 0}
    \definecolor{signalValue3}{RGB}{255, 243, 0}
    \definecolor{signalValue4}{RGB}{164, 164, 91}
    \definecolor{signalValue5}{RGB}{112, 112, 143}


    \begin{tikzpicture}
        
        \node[graphNode, fill=signalValue3] at (0, 1) (1)  {};
        \node[graphNode, fill=signalValue5] at (1, 0) (0)  {};
        \node[graphNode, fill=signalValue2] at (2, 1) (2)  {};
        \node[graphNode, fill=signalValue3] at (2, 2) (3)  {};
        \node[graphNode, fill=signalValue2] at (3, 0) (4)  {};
        \node[graphNode, fill=signalValue1] at (4, 1) (5)  {};
    
        \draw[graphEdge] (0) -- (2);
        \draw[graphEdge] (1) -- (2);
        \draw[graphEdge] (3) -- (2);
        \draw[graphEdge] (4) -- (2);
        \draw[graphEdge] (5) -- (2);
        
    \end{tikzpicture}

            \end{subFigure}
            \begin{subFigure}{0.2\linewidth}
                \colorbar{-1}{1}{2.5cm}
            \end{subFigure}
            \caption
            {
                Example of a signal on a star graph.
                The intensity of the signal on each node is given through the color bar.
            }
            \label{graphWithSignal}
        \end{figure}

        Another matrix that provides useful information on the graph is its normalized Laplacian \cite{Chung1997}.
        This is a differentiation operator that is analogous to the Laplacian that intervenes in the heat propagation equations or in harmonic analysis:
        
        \begin{definition}[Normalized Laplacian]
            The normalized Laplacian \NL{} associated with a graph \G{} with a matrix of weights \W{} is defined as $\NL \triangleq \I{\N} - \D^{-\frac{1}{2}} \W \D^{-\frac{1}{2}}$, where \D{} is a diagonal matrix of degrees (\ie{} $\forall \u \in \V : \matij{\D}{\u}{\u} = \sum\limits_{\v \in \V} \matij{\W}{\u}{\v}$), and \I{\N} is the $\N \times \N$ identity matrix.
        \end{definition}

        One property of the matrix \NL{} is that it is defined as a linear combination of real symmetric matrices.
        As a consequence, it is itself a real symmetric matrix, and can thus be diagonalized into an orthonormal basis.
        We denote the eigenvectors of this basis $\eigvec{\NL} = \set{\ev_1; \dots; \ev_\N}$, and the associated eigenvalues $\eigval{\NL} = \set{\el_1 \leq \el_2 \leq \dots \leq \el_\N}$.

        Shuman \etal{} have shown that there is a correspondence between the frequency of the Fourier basis in classical signal processing and the eigenvalues \eigval{\NL} of the normalized Laplacian.
        This property allowed researchers to define tools for signal processing on graphs such as graph Fourier transform, convolution of signals or wavelets on graphs \cite{Shuman2013}, among others.
        
        In this context of porting tools from classical signal processing to signal processing on graphs, Agaskar and Lu have shown \cite{Agaskar2012} that a signal on a graph could not be fully localized both in the graph domain and in the spectral domain.
        The details of their work is the object of the next section.

    %
    

    \section{The uncertainty principle applied to signal processing on graph}
    \label{uncertainty}
        
        Heisenberg's uncertainty principle states that a signal \x{} cannot be fully localized both in the time and frequency domains.
        More precisely, there exists an analytic expression quantifying this compromise:
            \begin{equation}
                \sqTS(\x) \sqFS(\x) \geq \frac{1}{4}
                \;,
                \label{heisenberg}
            \end{equation}
            where $\sqTS(\x)$ is the \emph{time spread} of the signal, and $\sqFS(\x)$ is its \emph{frequency spread}.
        
        In order to transpose this notion to signal processing on graphs, Agaskar and Lu \cite{Agaskar2012} propose to define notions that are analogous to $\sqTS(\x)$ and $\sqFS(\x)$.
        The graph being used to represent the support of the signals, thus generalizing the time line to more complex domains, finds its equivalent in a notion of \emph{graph spread}
            \footnote
            {
                We recall that we consider unit-norm signals.
                As a consequence, the \lTwo{} normalization in the original equation of \cite{Agaskar2012} is no longer required here.
            }:
        
        \begin{definition}[Graph spread]
            Let \x{} be a signal on a graph $\G = (\V, \E, \W)$.
            Let $\uc{} \in \V$.
            The graph spread $\sqGS(\x)$ of the signal around node \uc{} is defined by:
                \begin{equation}
                    \sqGS(\x) \triangleq \sum\limits_{\u \in \V} \dist(\uc, \u) \veci{\x}{\u}^2 = \x^\top \P \x
                    \;,
                    \label{graphSpread}
                \end{equation}
                where \dist{} is a distance function, and \P{} is the diagonal matrix of distances to node \uc.
                In the original definition \cite{Agaskar2012}, the authors use the squared geodesic distance for \dist.
                This choice has been discussed in \cite{Pasdeloup2015}.
        \end{definition}
        
        Moreover, the correspondence between the Fourier domain and the eigenvalues of \NL{} being established, Agaskar and Lu define a notion of \emph{spectral spread} \cite{Agaskar2012} for the signal:
        
        \begin{definition}[Spectral spread]
            Let \x{} be a signal on a graph $\G = (\V, \E, \W)$.
            The spectral spread $\sqSS(\x)$ of the signal is defined by:
                \begin{equation}
                    \sqSS(\x) \triangleq \sum\limits_{n = 1}^\N \el_n \veci{\sx}{n}^2 = \x^\top \NL \x
                    \label{spectralSpread}
                    \;,
                \end{equation}
                where $\set{\el_1; \dots; \el_\N}$ are the eigenvalues of \NL{}, and $\sx{} = \set{\sx_1; \dots; \sx_\N}$ is the graph Fourier transform \cite{Shuman2013} of \x{} on \G.
        \end{definition}
        
        Informally, these two notions quantify the concentration of a signal in the graph domain or in the spectral domain.
        As an example, consider a signal $\x = \set{1; 0; \dots; 0}$, with the only non-zero component being on node \uc.
        With this signal, we obtain $\sqGS(\x) = 0$, which corresponds to a signal that is fully localized on one node in the graph domain
        
        It is worth remarking that $\sqSS(\x)$ is minimized for a signal having all its energy on the first eigenvalue $\el_1 = 0$, which is also an observed property on signals diffused a high number of times on a non-bipartite graph.
        As a matter of fact, when a signal has all its energy on $\el_1$, then it is observable that it completely spread in the graph domain.
        
        Agaskar and Lu have shown that the compromise between localization of a signal in the graph domain and in the spectral domain is dependent on the topology of the graph.
        Therefore, to the best of our knowledge, no work has been done to provide a universal bound on the compromise (for a given \N), as it was classically made in \eqref{heisenberg}.
        Although, Rabbat and Gripon \cite{Rabbat2014} have shown that the minimal graph spread for a null spectral spread was obtained for the star graph, which is a first result in the obtention of such a bound.
        
        In order to study Heisenberg's uncertainty principle applied to signals on graphs, it is necessary to fix a graph \G{} and a node \uc{} used as reference for the graph spread.
        We can then determine an \emph{uncertainty curve} representing the best possible compromises:
        
        \begin{definition}[Uncertainty curve]
            The uncertainty curve \uCurve{} associated with a graph \G{}, for a chosen node \uc{}, is defined by:
                \begin{equation}
                    \uCurve(\alpha) \triangleq \min\limits_{\x} \sqGS(\x) ~\st~ \sqSS(\x) = \alpha
                    \label{uncertaintyCurve}
                    \;,
                \end{equation}
                and can be plotted by varying $\alpha$ from $0$ to $1$ \cite{Agaskar2012}.
        \end{definition}
        
        In the rest of this section, we choose to study an unweighted star graph, as in \figref{graphWithSignal}, and set \uc{} as the middle node of the graph.
        For this particular graph, the uncertainty curve is depicted in \figref{uncertaintyStarGraph}.
        Contrary to most graphs, it is possible to show that this curve is the same for every star graph, whatever the value of \N:
        
        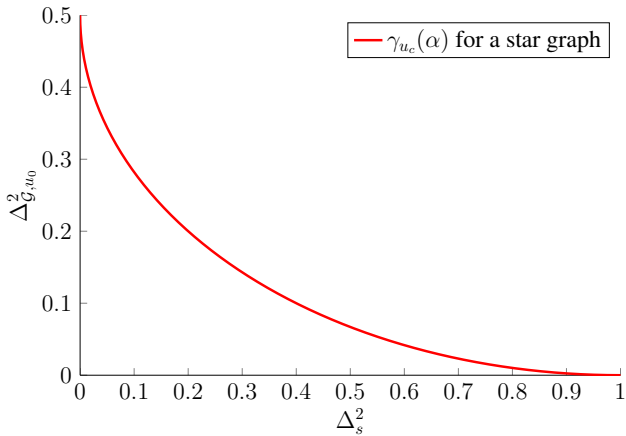
\begin{figure}[h!]
            \centering
            \Large
            {
                \resizebox{\hsize}{!}
                {
%
%
\begin{tikzpicture}

\begin{axis}[%
width=4.5in,
height=3in,
at={(0.808889in,0.513333in)},
scale only axis,
every outer x axis line/.append style={black},
every x tick label/.append style={font=\color{black}},
xmin=0,
xmax=1,
xlabel={$\Delta_s^2$},
every outer y axis line/.append style={black},
every y tick label/.append style={font=\color{black}},
ymin=0,
ymax=0.5,
ytick={0,0.1,0.2,0.3,0.4,0.5},
ylabel={$\Delta_{\mathcal{G}, u_0}^2$},
axis x line*=bottom,
axis y line*=left,
legend style={legend cell align=left,align=left,draw=black}
]

\addplot [line width=1.5pt, color=red, solid]
  table[row sep=crcr]{%
0	0.5\\
0.000301181303795762	0.487729385738535\\
0.00120454379482829	0.475466162836295\\
0.00270954332130741	0.463217718200188\\
0.00481527332780214	0.450991429835221\\
0.00752046540129404	0.438794662400375\\
0.01082349003522	0.42663476277232\\
0.0147223576110539	0.414519055619867\\
0.019214719596769	0.402454838991936\\
0.0242978699614722	0.390449379921566\\
0.0299687468054551	0.378509910048368\\
0.0362239342045604	0.366643621262551\\
0.0430596642677915	0.354857661372767\\
0.0504718194069693	0.343159129800547\\
0.0584559348169835	0.331555073303883\\
0.0670072011652691	0.320052481732494\\
0.076120467488713	0.308658283817456\\
0.0857902442964595	0.297379342997515\\
0.0960107068765509	0.286222453284865\\
0.10677569880448	0.275194335172703\\
0.11807873565164	0.264301631587006\\
0.129913008891263	0.25355090388513\\
0.142271389999719	0.242948627903395\\
0.155146434750303	0.232501190056445\\
0.168530387697452	0.2222148834902\\
0.182415186848426	0.212095904291068\\
0.196792468519368	0.202150347753774\\
0.211653572373422	0.192384204709669\\
0.226989546637267	0.182803357918176\\
0.242791153493496	0.173413578523123\\
0.259048874645033	0.164220522576495\\
0.275752917048526	0.155229727631471\\
0.292893218813452	0.146446609406726\\
0.310459455262912	0.137876458524277\\
0.328441045152976	0.129524437322522\\
0.346827157046236	0.121395576746752\\
0.365606715836355	0.113494773318631\\
0.384768409419392	0.10582678618669\\
0.404300695507572	0.0983962342596757\\
0.42419180858215	0.0912075934242097\\
0.4444297669804	0.0842651938487266\\
0.46500238011293	0.0775732173751386\\
0.485897255806791	0.0711356949998608\\
0.507101807770217	0.0649565044456434\\
0.528603263174009	0.059039367825821\\
0.550388670345374	0.0533878494022465\\
0.572444906569722	0.0480053534382771\\
0.594758685995039	0.0428951221482284\\
0.617316567634912	0.0380602337443566\\
0.640104963464983	0.0335036005826358\\
0.663110146607771	0.0292279674084909\\
0.686318259601115	0.0252359097034807\\
0.709715322745532	0.0215298321338966\\
0.733287242525074	0.0181119671022836\\
0.757019820096723	0.0149843734027293\\
0.780898759843123	0.0121489349807361\\
0.804909677983869	0.00960735979838488\\
0.829038111239712	0.00736117880552804\\
0.853269525544642	0.00541174501760944\\
0.877589324800774	0.00376023270064548\\
0.901982859670439	0.00240763666390154\\
0.926435436400333	0.00135477166065491\\
0.950932325672581	0.00060227189741379\\
0.975458771477089	0.000150590651897878\\
1	0\\
};
\addlegendentry{$\uCurve(\alpha)$ for a star graph};

\end{axis}
\end{tikzpicture}%
                }
                \caption
                {
                    Uncertainty curve associated with a star graph.
                    The middle node is chosen for \uc, and we use the squared geodesic distance for the computation of \sqGS{} in \eqref{graphSpread}.
                }
                \label{uncertaintyStarGraph}
            }
        \end{figure}
        
        The approach in \eqref{uncertaintyCurve} to obtain \uCurve{} requires to search the whole space of signals in order to find those that minimize \sqGS{} for a fixed value of \sqSS.
        We denote \~\x{} the signals that reach the uncertainty curve, \ie{} such as the pair $\pair{\sqGS(\~\x)}{\sqSS(\~\x)}$ represents a point in \uCurve.
        For a star graph, when choosing \uc{} as the middle node, Agaskar and Lu have shown that every \~\x{} is of the form $\~\x = \set{\~\x_1; \~\x_2; \dots; \~\x_2}$.
        Without loss of generality, it is possible to describe the signals reaching the uncertainty curve using the fact that they are included in a circle.
        Therefore, for a star graph with central \uc{}, \uCurve{} corresponds to the lower left portion of an ellipse of equation:
            \begin{equation}
                (\sqSS - 1)^2 + (2 \sqGS - 1)^2 = 1
                \;.
                \label{ellipseStar}
            \end{equation}
        
        For more complex graphs, the authors use an algorithm called the \emph{sandwich algorithm} to approximate the uncertainty curve with arbitrary precision.
        
    %
    

    \section{Reduction of the search space of signals reaching the uncertainty curve}
    \label{contribution}
        
        In order to reduce the search space for the resolution of \eqref{uncertaintyCurve}, we want to characterize the signals that reach the uncertainty curve.
        To propose such a characterization, we extend the work of Agaskar and Lu in \cite{Agaskar2012}, appendix C.
        This work was originally made to show that the solution signals \~\x{} for a complete graph or a star graph -- for \uc{} being the middle node -- have a particular form.
        The objective of this section is to generalize this approach to make it relevent for a larger class of graphs.
        We propose to prove \prref{generalization} :

        \begin{property}
            Let $\M(\alpha) \triangleq \P - \alpha \NL$ be a matrix defined for a fixed $\alpha$.
            If $\M(\alpha)$ is of the form:
                \begin{equation}
                    \M(\alpha) = \left(
                                     \begin{array}{c|c}
                                         \A & \B \\ \hline
                                         \C & \D
                                     \end{array}
                                 \right)
                    \label{mAlpha}
                    \;,
                \end{equation}
                where:
                \begin{itemize}
                    \item \A{} is a square matrix of dimension $j$.
                    \item \B{} is a matrix that is constant by line, \ie{} $\B = \y \vecOne{j}^\top$ for \y{} any vector, and for \vecOne{j} a vector of dimension $j$ with all components equal to $1$.
                    \item \C{} is any matrix.
                    \item \D{} is circulant of dimension $k$.
                \end{itemize}
                then \~\x{} is of the form:
                \begin{equation}
                    \set{\~\x_1; \dots; \~\x_j; \underbrace{\~\x_{j+1}; \dots; \~\x_{j+1}}_{k \text{ times}}}
                    \;.
                    \label{genericForm}
                \end{equation}
            \label{generalization}
        \end{property}
        
        For the needs of the proof, we recall \prref{xSolution} from \cite{Agaskar2012} :

        \begin{property}
            Every signal \~\x{} reaching the uncertainty curve is the eigenvector associated with the lowest eigenvalue of a matrix $\M(\alpha) \triangleq \P - \alpha \NL$.
            \label{xSolution}
        \end{property}

        \begin{proof}[Proof of \prref{generalization}]
            \item[$\bullet$] Since \D{} is circulant, we have that \vecOne{k} is an eigenvector for \D.
                             By construction of $\M(\alpha)$, \D{} is symmetric, and can thus be decomposed into an orthonormal basis.
                             Let $\set{\ev_1; \dots; \ev_{k-1}}$ be the eigenvectors of \D{} orthogonal to \vecOne{k}, associated with the eigenvalues $\set{\el_1; \dots; \el_{k-1}}$.
                             By construction, we have $\scalar{\ev_i}{\vecOne{k}} = 0, \forall i \in \intInterval{1}{k-1}$.
            \item[$\bullet$] For all $i \in \intInterval{1}{k-1}$, we build a vector $\ev_i^+ \triangleq \set{\vecZero{j}; \ev_i^\top}^\top$, where \vecZero{j} is the null vector of dimension $j$.
                             Using the fact that \B{} is constant by line, we obtain that $\forall i : \M(\alpha) \ev_i^+ = \el_i \ev_i^+$.
                             Therefore, every $\ev_i^+$ is eigenvector of $\M(\alpha)$.
                             Using the methodology of \cite{Agaskar2012}, appendix C (Rayleigh inequality), we obtain that the eigenvector associated with the smallest eigenvalue of $\M(\alpha)$ must be orthogonal to $\ev_i^+, \forall i \in \intInterval{1}{k-1}$.
            \item[$\bullet$] By noting that the $j$ first components of vectors $\ev_i^+$ are null, and by application of \prref{xSolution}, we obtain that every vector \~\x{} attaining the uncertainty curve is of the form in \eqref{genericForm}.
        \end{proof}
        
        It is interesting to remark that the application of \prref{generalization} can be made recursively on \A, allowing one to refine the characterization of the $\~\x_1 \dots \~\x_j$ components in \eqref{genericForm}, and thus to reduce the search space of solutions.
        
        To illustrate the characterization of signals \~\x{} reaching the uncertainty curve, let us consider a star graph, but this time with \uc{} taken as one of the \emph{leave nodes} of the graph.
        We obtain that $\M(\alpha)$ can be decomposed in a way that a square circulant submatrix \D{} of dimension $\N-2$ appears.
        Using \prref{generalization}, we obtain that, for this graph and this choice of \uc{}, all signals \~\x{} are of the form $\set{\~\x_1; \~\x_2;\~\x_3; \dots; \~\x_3}$.
        
        In order to iterate over the solution signals \~\x, we can consider the set of unit-norm signals defined on an hypersphere of dimension $M$, where $M$ is the number of distinct components in the characterization of \~\x.
        In the considered example, we can thus reduce the set of potential solutions to signals characterized by two parameters $\theta$ and $\phi$ as follows: $\set{\cos \theta; \sin \theta \cos \phi; \frac{\sin \theta \sin \phi}{\sqrt{N - 2}}}$.
        
\begin{figure}[h!]
            \centering
            \includegraphics[width=\linewidth]{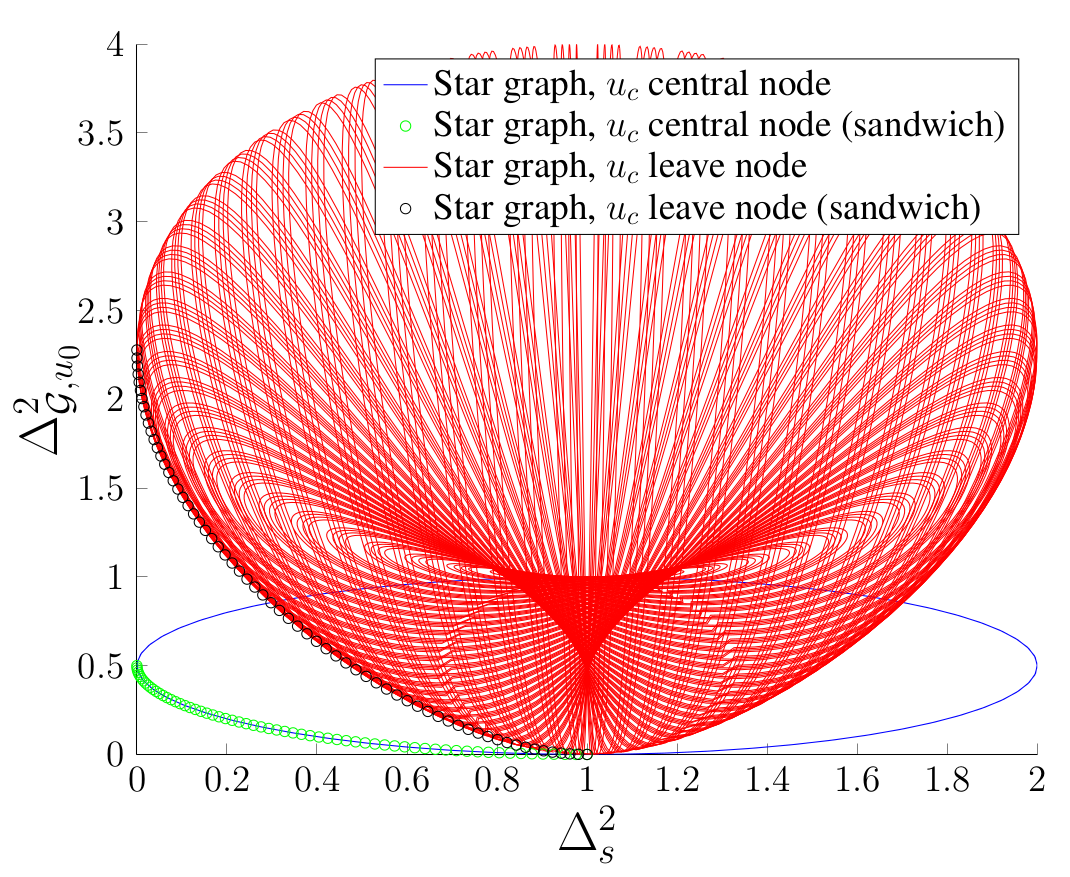}
            \caption
            {
                Sub-sampling of the space of potential solutions for a star graph, for different choices of \uc.
                The lower left border of the set of explored signals is the uncertainty curve.
                Approximation through the sandwich algorithm matches the obtained results.
            }
            \label{ellipses}
        \end{figure}

        \figref{ellipses} represents the pairs $\pair{\sqGS(\~\x)}{\sqSS(\~\x)}$ for signals \~\x{} obtained by sub-sampling, by iterating over the possible values of $\theta$ and $\phi$ in the interval $\interval{0}{2\pi}$, with a step of $0.05$.
        We also depict the uncertainty curves obtained using the sandwich algorithm, and observe that they match the frontier of the set of explored signals.

    %
    

    \section{Conclusion}
    \label{conclu}
        
        In this document, we have presented an extension of the method introduced by Agaskar and Lu in \cite{Agaskar2012} in order to characterize the signals that realize the uncertainty curve, for a given graph and a chosen reference node \uc.
        Whereas the original approach was only developped for complete graphs and star graphs (only when \uc{} was chosen as the middle node), our method can give a characterization of solution signals for a larger class of graphs.
        To illustrate our results, we have plotted a subset of potential solution signals in the case of a star graph when choosing \uc as a leaf, and have observed that our results are matched by those provided by the approached sandwich algorithm.
        
        To improve our work, we will first try to find a method to reduce the size of the search space to discriminate signals on the uncertainty curve from other potential solutions.
        Also, we will study in more details the impact of the choice of \uc, in order to be able to propose a \emph{canonical} uncertainty curve and to allow comparison of different graphs.
        Finally, another direction of our work will be to try to determine, for a fixed \N, what could be a universal uncertainty curve for every possible graph topology, thus enabling to state an uncertainty principle similar to \eqref{heisenberg}.
        
    %
    
    
    \vfill\pagebreak
    \label{refs}
        
        \bibliographystyle{IEEEbib}
        \bibliography{paper}

\begin{thebibliography}{1}

\bibitem{Shuman2013}
David~I. Shuman, Sunil~K. Narang, Pascal Frossard, Antonio Ortega, and Pierre
  Vandergheynst,
\newblock ``Signal processing on graphs: Extending high-dimensional data
  analysis to networks and other irregular data domains,''
\newblock {\em CoRR}, vol. abs/1211.0053, 2012.

\bibitem{Agaskar2012}
Ameya Agaskar and Yue~M. Lu,
\newblock ``A spectral graph uncertainty principle,''
\newblock {\em CoRR}, vol. abs/1206.6356, 2012.

\bibitem{Agaskar2012b}
Ameya Agaskar and Yue~M Lu,
\newblock ``Uncertainty principles for signals defined on graphs: Bounds and
  characterizations,''
\newblock in {\em Acoustics, Speech and Signal Processing (ICASSP), 2012 IEEE
  International Conference on}. IEEE, 2012, pp. 3493--3496.

\bibitem{Folland1997}
GeraldB. Folland and Alladi Sitaram,
\newblock ``The uncertainty principle: A mathematical survey,''
\newblock {\em Journal of Fourier Analysis and Applications}, vol. 3, no. 3,
  pp. 207--238, 1997.

\bibitem{Tsitsvero2015}
Mikhail Tsitsvero, Sergio Barbarossa, and Paolo Di~Lorenzo,
\newblock ``Signals on graphs: Uncertainty principle and sampling,''
\newblock {\em arXiv preprint arXiv:1507.08822}, 2015.

\bibitem{Chung1997}
Fan~RK Chung,
\newblock {\em Spectral graph theory}, vol.~92,
\newblock American Mathematical Soc., 1997.

\bibitem{Pasdeloup2015}
Bastien Pasdeloup, R{\'{e}}da Alami, Vincent Gripon, and Michael Rabbat,
\newblock ``Toward an uncertainty principle for weighted graphs,''
\newblock {\em CoRR}, vol. abs/1503.03291, 2015.

\bibitem{Rabbat2014}
Michael Rabbat and Vincent Gripon,
\newblock ``Towards a spectral characterization of signals supported on
  small-world networks,''
\newblock in {\em ICASSP 2014 : IEEE International Conferences on Acoustics,
  Speech and Signal Processing}, IEEE, Ed., 2014.

\end{thebibliography}
        
    %
    

\end{document}